\documentclass[]{pasj02} 
\usepackage[switch,mathlines]{lineno} % add line number to manuscript

\usepackage{rotating}
\usepackage{comment}
\jyear{2024}
\Received{}%{yyyy/mm/dd}
\Accepted{}%{yyyy/mm/dd}
\begin{document}

\title{\textcolor{red}{A} High Incidence of Mid-infrared Variability in \\ Local Ultraluminous Infrared Galaxies }

%%% begin:list of authors
% Do NOT capitalize all letters in "textsc".
\author{
 Shun \textsc{Hatano,}\altaffilmark{1,2}\altemailmark\orcid{0000-0002-5816-4660} \email{shun.hatano@grad.nao.ac.jp}
  Masatoshi \textsc{Imanishi,}\altaffilmark{1,2}\orcid{0000-0001-6186-8792}
  Takanobu \textsc{Kirihara,}\altaffilmark{3}\orcid{0000-0001-6503-8315}
  Takashi  \textsc{Yamamoto,}\altaffilmark{4}\orcid{}
  Yuxing \textsc{Zhong,}\altaffilmark{4}\orcid{0009-0001-3910-2288}
  and Chenghao \textsc{Zhu}\altaffilmark{5,6}\orcid{0000-0002-9888-6895}
  %B-Firstname \textsc{B-Familyname},\altaffilmark{2}$^{,\dag}$\orcid{0000-0000-0000-0000}
 %C-Firstname \textsc{C-Familyname},\altaffilmark{3}\altemailmark \email{ccccc@xxx.xxx.xx.xx}
 %and 
 %D-Firstname \textsc{D-Familyname}\altaffilmark{2}\altemailmark\orcid{0000-0000-0000-0000} \email{ddddd@xxx.xxx.xx.xx}
}
\altaffiltext{1}{National Astronomical Observatory of Japan, Osawa 2-21-1, Mitaka, Tokyo 181-8588, Japan}
\altaffiltext{2}{Department of Astronomical Science, The Graduate University for Advanced Studies, SOKENDAI, 2-21-1 Osawa, Mitaka, Tokyo, 181-8588, Japan}
\altaffiltext{3}{Kitami Institute of Technology, 165, Koen-cho, Kitami, Hokkaido 090-8507, Japan}
\altaffiltext{4}{Department of Physics, School of Advanced Science and Engineering,
Faculty of Science and Engineering, Waseda University, 3-4-1, Okubo,
Shinjuku, Tokyo 169-8555, Japan}
\altaffiltext{5}{Institute for Cosmic Ray Research, The University of Tokyo, 5-1-5 Kashiwanoha, Kashiwa, Chiba 277-8582, Japan}
\altaffiltext{6}{Department of Physics, Graduate School of Science, The University of Tokyo, 7-3-1 Hongo, Bunkyo, Tokyo 113-0033, Japan}
%\altaffiltext{3}{C-Address of Institute}

%\footnotetext[$\dag$]{}

%%% end:list of authors

%% !!! Select 3 to 5 words from PASJ's key words !!! 
%% List of Key Words: https://academic.oup.com/pasj/pages/Pasj_Keywords 
%% "\KeyWords{ }" always has to be placed before ``\maketitle'' 

\maketitle

\begin{abstract}

We explore mid-infrared (MIR) variability in local ultraluminous infrared galaxies ({ULIRGs; infrared {luminosity} $L_{\rm IR}>10^{12}\ L_\odot$}) utilizing the $\sim$~11~years of photometry from the NEOWISE multi-epoch catalog of {\it Wide-field Infrared Survey Explorer} ({\it WISE}). We identify 30 variable ULIRGs with statistically significant MIR variability. The variability is observed on timescales of a few years, implying that the {variable} MIR-emitting regions are compact {($\lesssim \mathrm{a \ few\ pc}$)}. The difference between maximum and minimum $W2$ (4.6 ${\rm \mu}$m) band luminosity ($\Delta L_{\rm W2}$) of the 30 variable ULIRGs {ranges} from $\Delta L_{W2}$ = {$7\times10^{42}$} to $5\times 10^{44}$ erg s$^{-1}$. The $\Delta L_{W2}$ of 25 variable ULIRGs out of 30 are greater than $\Delta L_{W2}$ = $1\times10^{43}$ erg s$^{-1}$, surpassing the MIR luminosity {range} observed in known supernovae (SNe; $L_{\rm 3.6\ {\rm \mu m}}$ and $L_{\rm 4.5\ {\rm \mu m}}$ < 10$^{42.3}$  erg s$^{-1}$). {Therefore, the MIR variabilities in these 25 ULIRGs are most likely driven by tidal disruption events (TDEs) or active galactic nuclei (AGN) {activities}.  Our sample {includes} hard X-ray detected AGNs (e.g., UGC 05101) and previously reported TDE candidates (IRAS F01004-2237, IRAS 05189-2524). All 25 also exhibit at least one AGN signature(s) beside{s} the MIR variability, suggesting that even if the MIR variability originate{{s}} from TDEs, the black hole{s} responsible are likely AGNs. 
{Our results suggest that MIR variability is an effective tool for detecting buried AGNs and highlights intense nuclear activity in ULIRGs. 
}}

\end{abstract}

%\pagewiselinenumbers 

\section{Introduction}

Almost all massive galaxies host a supermassive black hole (SMBH) at their center, and SMBH mass correlates with bulge stellar mass (e.g., \cite{1998AJ....115.2285M,2013ARA&A..51..511K}). 
ULIRGs, which could result from gas-rich mergers (\cite{2006asup.book..285L}), are {a} promising site of bulge and SMBH growth (\cite{1996ARA&A..34..749S, 2008ApJS..175..356H}). 
However, in ULIRGs, the growing SMBH is often deeply embedded in dust, blocking ionizing photons from reaching the narrow-line region and preventing optical diagnostics such as the Baldwin–Phillips–Terlevich (BPT; \cite{1981PASP...93....5B}) diagram from detecting these buried AGNs (\cite{2003MNRAS.344L..59M, 2007ApJS..171...72I,2008PASJ...60S.489I}). 
Therefore, observations at wavelengths less affected by {dust} extinction 
are necessary. 

{MIR observations, which are less affected by {extinction} than optical wavelengths (\cite{2008ApJ...680.1174N,2009ApJ...696.1407N,2011ApJ...737...73F}), provide a useful tool to investigate AGNs in ULIRGs.
\cite{2012ApJ...753...30S} showed that {the MIR color $W1 - W2$ ([3.4]-[4.6]) {{is}} an effective indicator for distinguishing AGNs from normal galaxies, as the hot dust emission heated by AGNs exhibits systematically redder MIR colors than those of the stellar emission from galaxies.}
However, the red color {seen in galaxies} can be mimicked by dust emission from  {star-forming} regions in particular cases (e.g., \cite{2016ApJ...832..119H,2018ApJ...858...38S}).
This limitation motivates the use of MIR variability as an additional diagnostic.
Variability provides a clear distinction between AGN activity and star formation.
{AGNs are known to vary across the electromagnetic spectrum (e.g., \cite{2004ApJ...601..692V,2016ApJ...826..118K}).
{In the infrared, variable emission originates from hot dust on sub-parsec to parsec scales, which reprocesses UV/optical emission from the central accretion disk, as indicated by time lags between optical and infrared emission (e.g.,  \cite{2006ApJ...639...46S,2014ApJ...788..159K,2014ApJ...788...48S}).  
}
Star formation, by contrast, typically extends over kiloparsec scales and is therefore not expected to vary on observable timescales, except for SNe.}
However, even the brightest SNe detected in the MIR have $Spitzer$ 4.5 ${\rm}$${\rm \mu}$m-band luminosities only up to $10^{42.3}$ erg s$^{-1}$ (\cite{2019ApJS..241...38S}).
Therefore, applying a luminosity threshold can effectively remove contamination from bright SNe.}
{In some cases, there are  very compact ($\sim$100 pc) nuclear starbursts observed in nearby ULIRGs (e.g., \cite{2000AJ....119..509S,2010ApJ...723..993D, 2011AJ....141..156I, 2017ApJ...843..117B,2021A&A...651A..42P}). 
However, the light-crossing time would still exceed a few $\times 10^{2}$ yr, an order of magnitude longer than the timescales probed by \textit{WISE}.}

{The variability observed in AGNs is generally attributed to changes in the accretion disk (e.g., \cite{1997ARA&A..35..445U,2018ApJ...864...27S}).
{Changes} in the accretion disk can arise from instabilities in the accretion disk or from inhomogeneous temperature fluctuations (\cite{1976MNRAS.175..613S}; \cite{2009ApJ...698..895K}; \cite{2014ApJ...783..105R}), or extreme enhancement of the accretion rate  which is often discussed in changing-look AGNs (CLAGNs; \cite{2016ApJ...826..188R,2016MNRAS.457..389M}). Increase of accretion rate may originate from TDEs \citep{1988Natur.333..523R,2021ARA&A..59...21G}.
TDEs are triggered when a star enters within the tidal radius of an SMBH, leading to the star’s disruption and subsequent accretion of debris. 
CLAGNs often exhibit large MIR variability (variability amplitude larger than 0.4 mag; \cite{2017ApJ...846L...7S,2018ApJ...864...27S,2018ApJ...866...26A,2020ApJ...889...46S}), 
and TDEs (and their candidates) show IR echoes, which are reprocessed emission from dust surrounding the central black hole
(theoretically {calculated} 
by \cite{2016MNRAS.458..575L} and observed by \cite{2016ApJ...829...19V,2016ApJ...828L..14J,2016ApJ...832..188D,2017ApJ...850...63J,2017ApJ...841L...8D,2019ApJ...885..110Y,2019ApJ...871...15J, 2020ApJ...898..129S, 2021ApJS..252...32J,2021ApJ...911...31J,2022ApJS..258...21W,2022ApJ...930L...4W, 2022A&A...664A.158R, 2023MNRAS.525.4057H, 2024ApJ...961..211M,2025A&A...704A...3K}).}

{
TDE candidates are found in ULIRGs (\cite{2017NatAs...1E..61T,2024A&A...692A.262S,2022A&A...664A.158R}) and IR {echoes} are also detected in MIR (\cite{2017ApJ...841L...8D,2022A&A...664A.158R}). 
The MIR luminosity of some TDE candidates even exceeds the most luminous SNe observed in MIR (\cite{2022A&A...664A.158R}).
 While {the} majority of TDEs occur in galaxies without an AGN (\cite{1988Natur.333..523R,2021ARA&A..59...21G,2023ApJ...948L...5P,2024ApJ...961..211M}), TDE candidates reported in ULIRGs are discussed to host AGNs (\cite{2017NatAs...1E..61T}; \cite{2022A&A...664A.158R}).
}

{Part of the systematic MIR variability analysis of ULIRGs was conducted by \cite{2022A&A...664A.158R}. However, they mainly focused on identifying TDE candidates in luminous infrared galaxies (LIRGs; $L_{\rm IR} > 10^{11}\ L_\odot$) and ULIRGs, and {were} not aimed at detecting variability originating from AGNs.}
In addition, their ULIRG sample was restricted to galaxies listed in \cite{2003AJ....126.1607S}. While \cite{2003AJ....126.1607S} includes more LIRGs, later published \cite{2010MNRAS.405.2505N} contains more ULIRGs.

{In this work, we apply statistical variability detection techniques that were originally developed for dwarf galaxies (e.g., \cite{2020ApJ...900...56S,2022ApJ...936..104W,2023ApJ...945..157H,2024ApJ...975...60A,2026ApJ...997..241H}) to the larger and more complete ULIRG catalog of \cite{2010MNRAS.405.2505N}.

{{We note that variability has another advantage. While imaging observations have suggested that the hot dust in ULIRGs is confined within compact regions of $\lesssim$100 pc (e.g., \cite{2000AJ....119..509S}), the typical spatial scale of the hot dust has remained uncertain. Variability studies can provide much tighter constraints, potentially limiting the MIR-emitting region to a few parsecs or less.}}

{We {note} that \cite{2022MNRAS.512.5183E} and \cite{2018ApJ...858...59M} investigate the AGN contribution in {LIRGs} via detailed SED fitting, and \cite{2025ApJ...979...68D} focus on outflows observed in ULIRGs with JWST, suggesting the contribution of AGNs in (U)LIRGs. Variability will offer independent verification for these studies.}

The structure of this paper is as follows: In Section 2, we explain the ULIRG sample. 
In Section 3, we outline the methodology for creating light curves and detecting time variability.
Section 4 presents the results. 
In Section 5, we {discuss} the origin of the variability. 
Finally, Section 6 provides a summary of our findings.
{Throughout this paper, magnitudes are based on the Vega system, and we adopt $H_0 = 70$ km s$^{-1}$ Mpc$^{-1}$,  $\Omega_{\rm M}=0.30$, and $\Omega_{\rm \Lambda} = 0.70$. }

\section{Sample and Data}\label{sec:2}

We select all 164 ULIRGs in \cite{2010MNRAS.405.2505N} which summarizes local bright ULIRGs that are spectroscopically observed with $Spitzer$ satellite. 
{{Their} sample is based on the {\it Infrared Astronomical Satellite (IRAS}}) Point Source Catalog Redshift (PSCz) survey \citep{2000MNRAS.317...55S} and the 1 Jy ULIRG sample at 60 ${\rm \mu}$m \citep{1998ApJS..119...41K}, and covers the full local ULIRG luminosity range without strong bias toward or against AGN activity.}
\cite{2010MNRAS.405.2505N} also summarizes redshift, optical type, and AGN contribution evaluated from spectroscopic data for all sources, which faciliates the comparison between MIR variability and other {independent} AGN criteria in our study. We show redshift and $W2$ band {luminosity} distribution of our sample  in figure 1. {The $W2$ band fluxes are} obtained from AllWISE source catalog (\cite{2010AJ....140.1868W}). 
{The {\it WISE} magnitudes are converted into fluxes from Vega magnitude system with zero-point fluxes of $f_{\nu,0}$ =309.540 and 171.787 Jy for the $W1$ and $W2$ bands, respectively.}

\begin{figure}
 \begin{center}
  \includegraphics[width=9 cm]{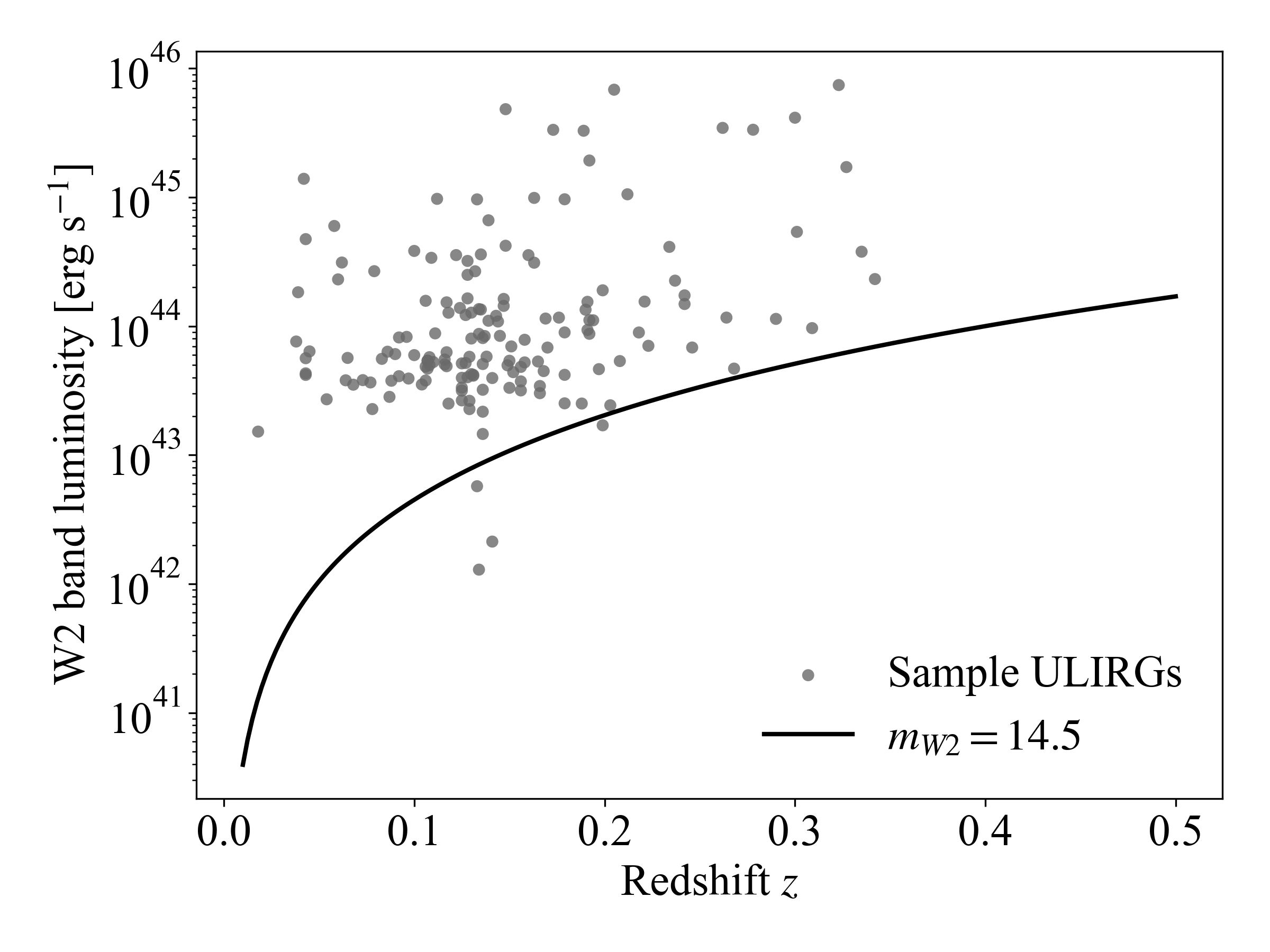}
 \end{center}
\caption{Redshift distribution and $W2$ band luminosity of {the} sample ULIRGs.
 {The black line represents the $W2$ luminosity corresponding to a source with $W2 = 14.5$ mag at each redshift. Most of our sample {is} {distributed} above this line. 
 {Alt text: Scatter plot with redshift on the horizontal axis and W2 band luminosity on the vertical axis, showing the sample of ultraluminous infrared galaxies. A diagonal reference line indicates the luminosity for magnitude fourteen point five, and most points appear above it.}
 } 
}\label{fig:sample}
\end{figure}

\begin{figure}
 \begin{center}
   \includegraphics[width=9cm]{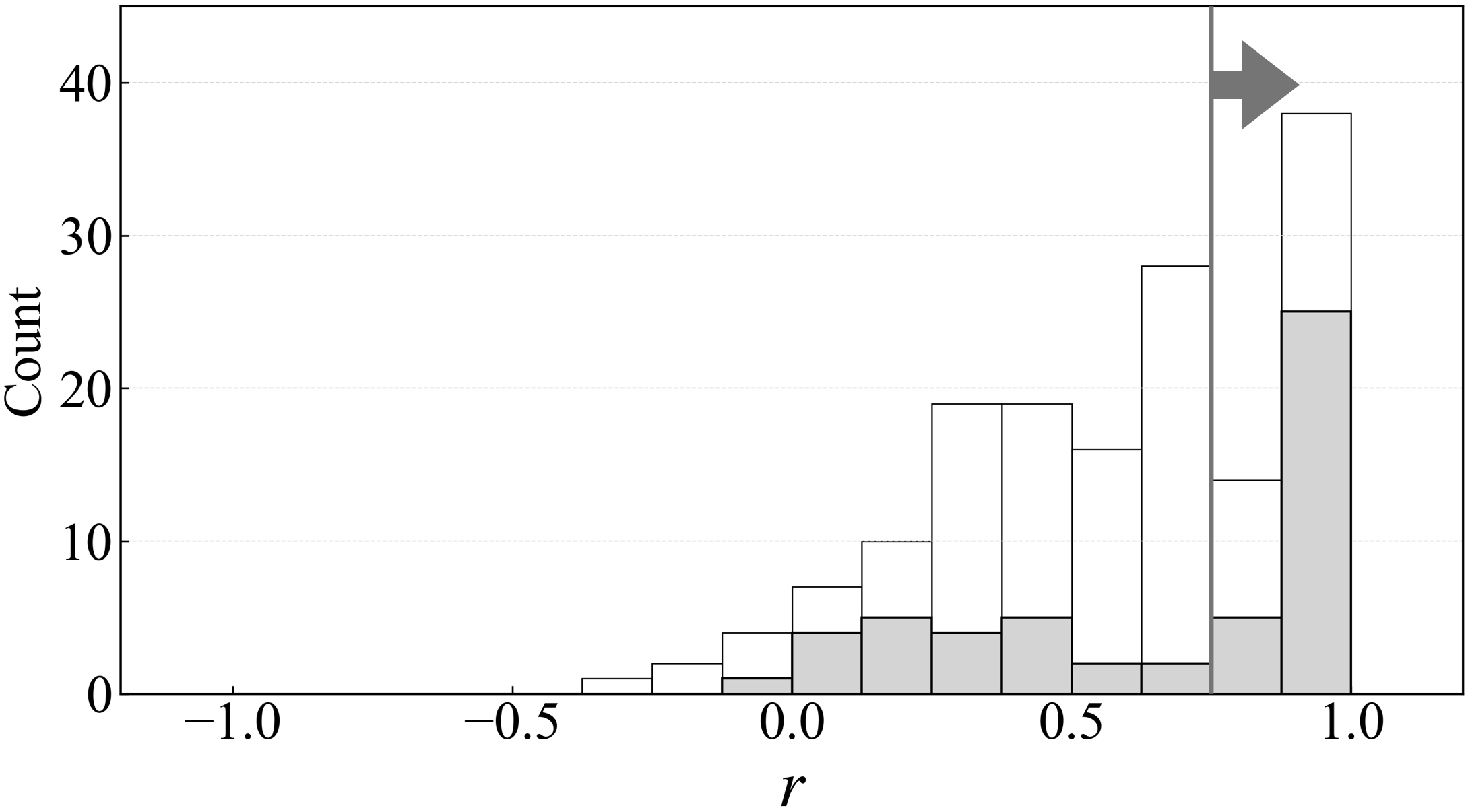} 
 \end{center}
\caption{{Distribution of the Pearson correlation coefficient $r$ for our sample with $W2 < 14.5$, $\sigma_{W1}>0.01$, and $\sigma_{W2}>0.034$. The gray vertical solid line and the arrow represent the selection criterion of $r>0.75$. {The white histogram shows the $r$ distribution of all sources with $W2 < 14.5$ before applying the additional criteria.}
Alt text: Histogram of Pearson correlation coefficient showing many objects above 0.75, indicating a large fraction of variable sources.
}
 { }%I would limit to $\sigma_{W1}>0.024$, $\sigma_{W2}>0.028$ sample. } 
}\label{fig:sample}
\end{figure}

\begin{figure}
 \begin{center}
  \includegraphics[width=9cm]{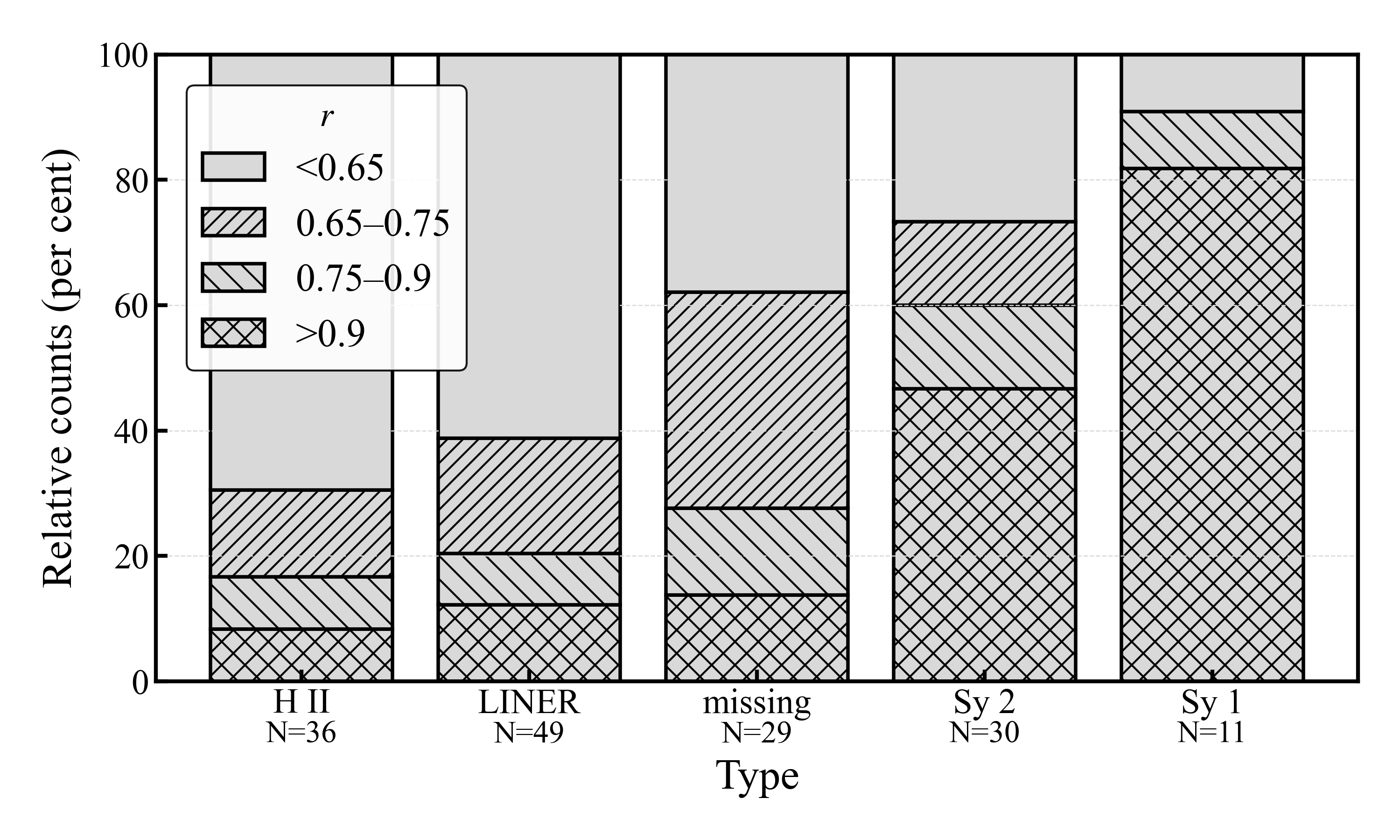} 
 \end{center}
\caption{Comparison of optical classifications with $r$ values. The number of entries is indicated below each bar. 
{Alt text: Bar chart showing Seyfert types have higher correlation r values, indicating stronger variability than other types.}
}\label{fig:sample}
\end{figure}

\begin{figure*}
 \begin{center}
  \includegraphics[width=15cm]{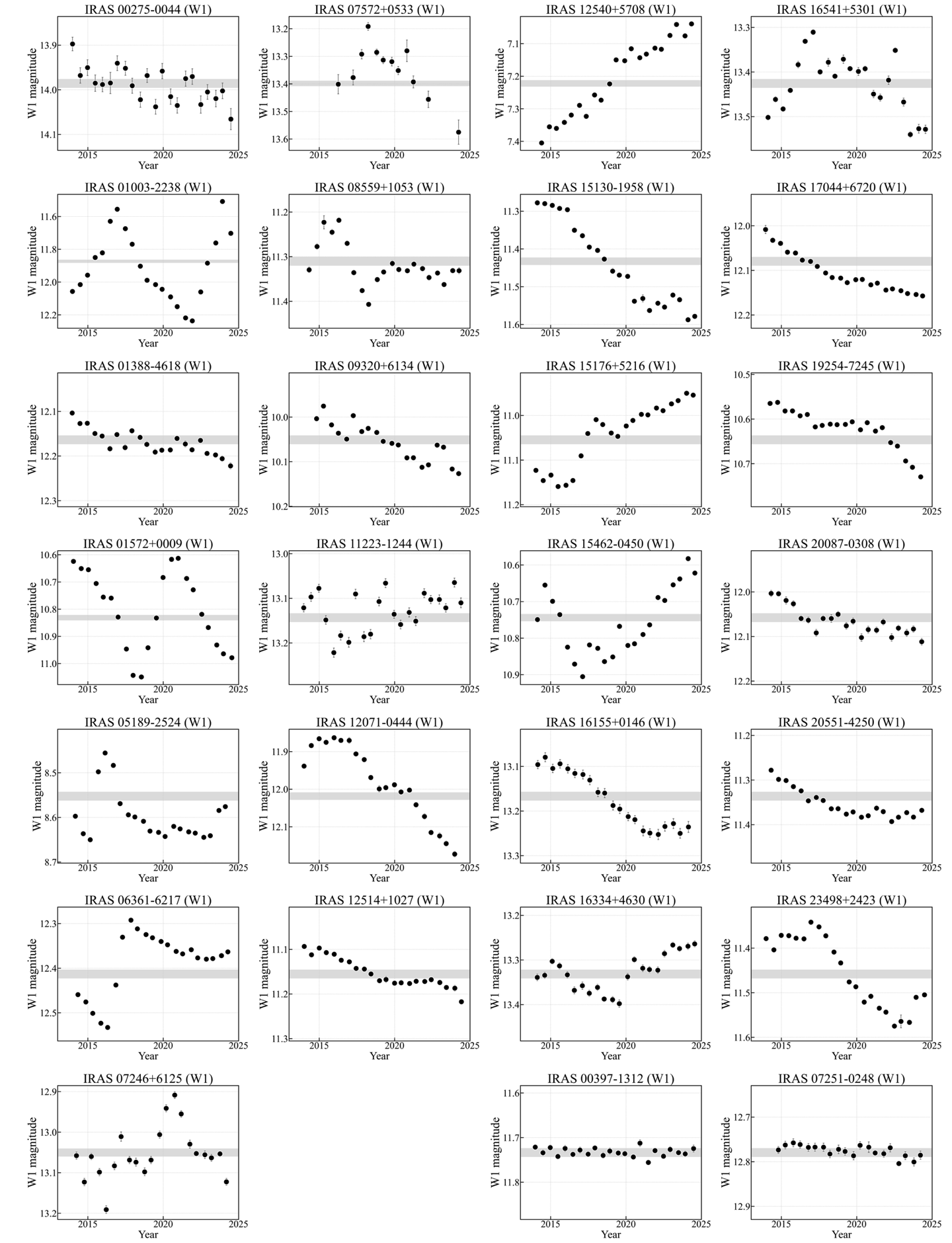} 
 \end{center}
\caption{{Binned $W1$ band light curves of variable ULIRGs with $L_{\rm W2}>10^{43}$ erg s$^{-1}$ (black circles). {We plot two ULIRGs at the lower right that did not meet our variability criteria. The gray horizontal band indicates twice the systematic uncertainty in $W1$ band. }
{Alt text: Light curves over an eleven year baseline. The vertical axis shows magnitude and the horizontal axis shows time. All variations are smaller than 0.7 magnitude but exceed the systematic width of 0.020 magnitude.}
}
}\label{fig:sample}
\end{figure*}

\begin{figure*}
 \begin{center}
  \includegraphics[width=15cm]{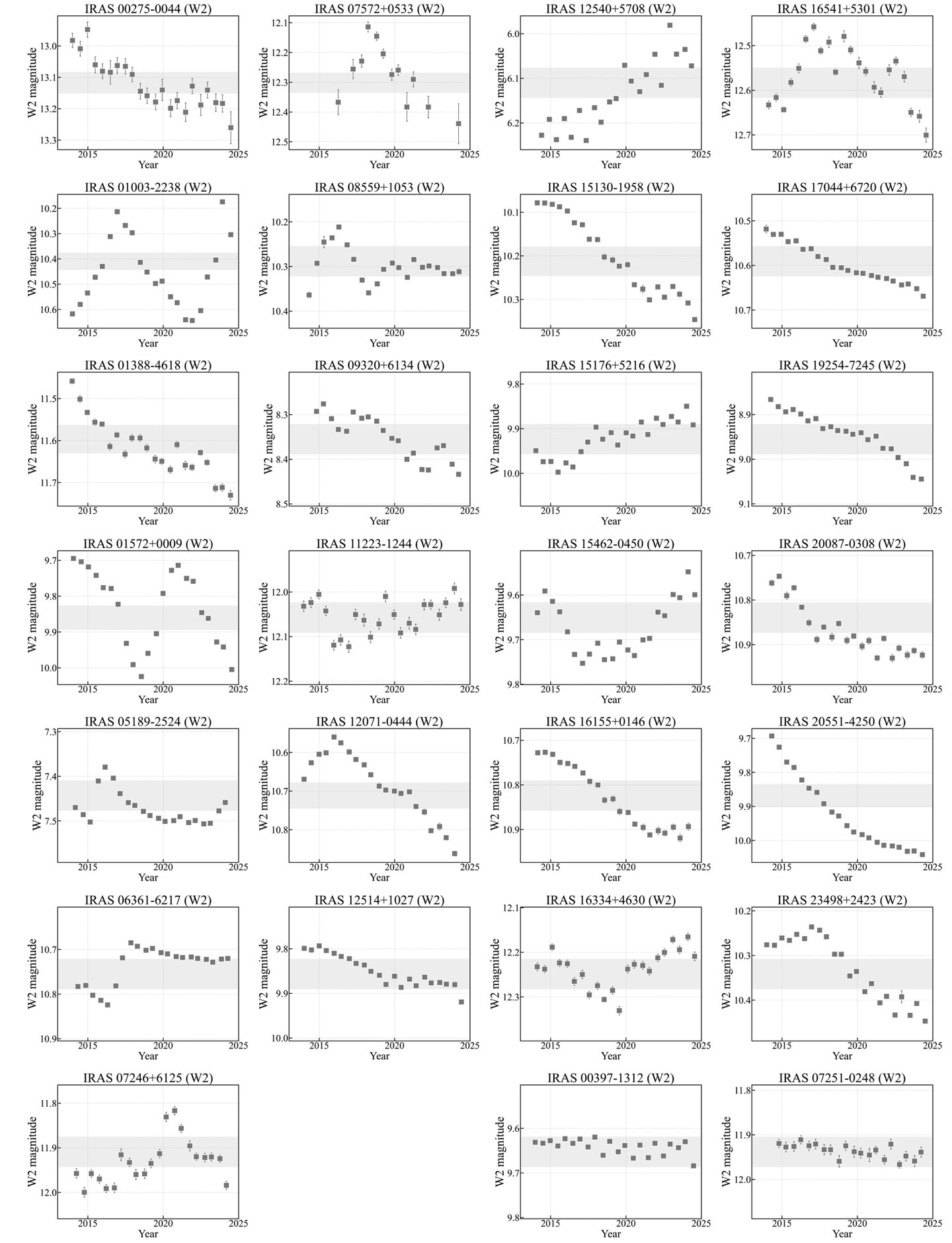} 
 \end{center}
\caption{{Binned $W2$ band light curves of variable ULIRGs with $L_{\rm W2}>10^{43}$ erg s$^{-1}$ (gray squares). {We plot two ULIRGs at the lower right that did not meet our variability criteria. The gray horizontal band indicates twice the systematic uncertainty in $W2$ band. }
{Alt text: Light curves over an eleven year baseline. The vertical axis shows magnitude and the horizontal axis shows time. All variations are smaller than 0.7 magnitude but exceed the systematic width of 0.068 magnitude.}
}
}\label{fig:sample}
\end{figure*}

\begin{figure*}
 \begin{center}
  \includegraphics[width=18cm]{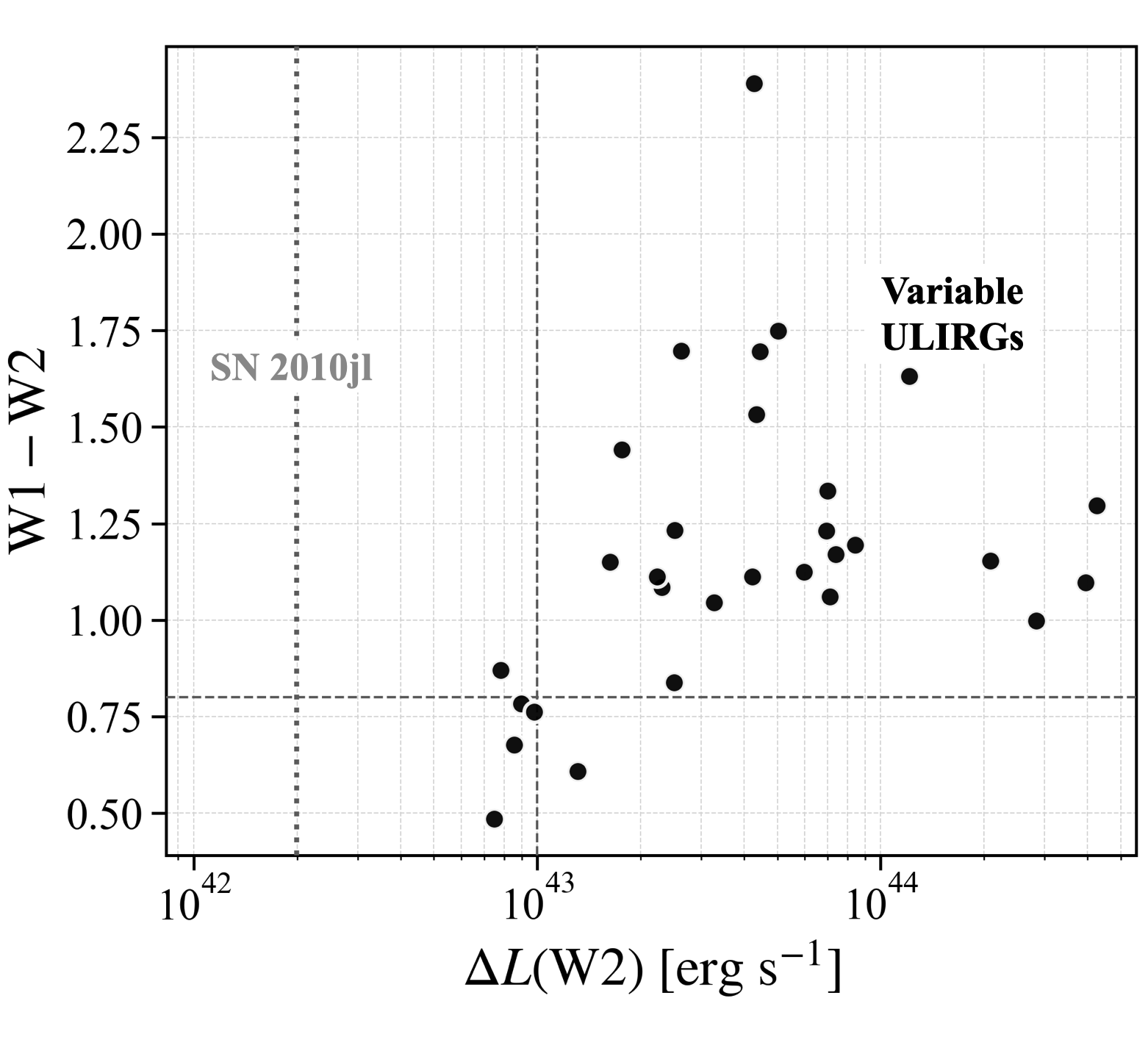} 
 \end{center}
\caption{{Distribution of $\Delta L_{W2}$ as a function of $W1 - W2$ color for variable ULIRGs (black circles). The horizontal {dashed} line represents the AGN selection criterion from \cite{2012ApJ...753...30S} ($W1 - W2 > 0.8$). The vertical {dashed} line indicates $\Delta L_{W2} = 10^{43}$ erg s$^{-1}$, which corresponds to approximately five times the  $\Delta L_{W2}$ of the most luminous supernova observed in MIR (SN 2010jl; the vertical dotted line). 
{Alt text: Scatter plot of W1 minus W2 color from 0.5 to 2.5 and delta W2 luminosity from ten to the forty two to five times ten to the forty four erg per second. Many galaxies lie above both the 0.8 color threshold and luminosities greater than ten to the forty three.}
}
}\label{fig:sample}
\end{figure*}

We use MIR, multi-epoch photometric data from the Wide-field Infrared Survey Explorer (WISE; \cite{2010AJ....140.1868W}). The mission was first carried out as the AllWISE  program mapping sky in four bands: $W1$ (3.4 ${\rm \mu}$m), $W2$ (4.6 ${\rm \mu}$m), $W3$ (12 ${\rm \mu}$m), and $W4$ (22 ${\rm \mu}$m) in 2010. 
After cryogen depletion, the mission survey continued post-cryogenic NEOWISE {Reactivation mission} (\cite{2014ApJ...792...30M}). Notably, NEOWISE Reactivation mission monitored the entire sky in the $W1$ and $W2$ bands for approximately 11 years (2013–2024).

Using the {\it IRAS} names compiled by \cite{2010MNRAS.405.2505N}, we obtain J2000 right ascension and declination coordinates with Astropy (\cite{2022ApJ...935..167A}). At each position, we cross-match to the AllWISE Source Catalog by selecting the nearest detection within {default} 10 arcsec and find that {163} out of 164 ULIRGs have secure AllWISE counterparts. For those matched ULIRGs, we extract single-exposure photometry from the NEOWISE-R Single Exposure (L1b) Source Table by retrieving all detections whose reported positions lie within 3 arcsec of the AllWISE {coordinate.}

\section{Method}\label{sec:4}

In this section, we explain our variability analysis. 
We apply a cut of $W2 < 14.5$ mag proposed by \cite{2020ApJ...900...56S}, which effectively removes the Eddington bias from the light curves.
Then, we 
{apply quality cuts to the single-epoch photometry}
%impose quality cut against single epoch measurements for single-epoch phtometry
with ${\rm qual\_flag} \geq 5$, and ${\rm moon\_masked=0}$. 
The NEOWISE single-epoch photometry consists of ‘epochs’ taken roughly every six months, and each target has about 22 epochs in total. 
Because we focus on {half-year} or longer {time} scale variability, we bin photometry for each epoch by deriving the weighted average and standard errors and {obtain the} binned flux and flux errors for each epoch, while conducting {three-sigma} clipping (we adopt {three} times median absolute deviation) at the same time to remove {{outliers}}. %the outlier. 
{This procedure yields $m_{W1,i}$ ($m_{W2,i}$): the $W1$- ($W2$-) band photometric magnitude measured at the $i$-th epoch.}
We use {the} binned {photometric} measurements and its errors to calculate {{statistics}} for each object. 
We calculate unbiased Pearson $r$ correlation coefficients:
\begin{equation}
    r = \frac{C_{W1, W2}}{\sigma_{W1} \sigma_{W2}},
\end{equation}
where $C_{W1, W2}$ denotes the unbiased covariance between $W1$- and $W2$- band photometry {of} individual sources, 
\begin{equation}
    C_{W1, W2} = \frac{1}{N-1} \sum_i^{N} (m_{W1,i} - \langle m_{W1}\rangle ) \times (m_{W2,i} - \langle m_{W2}\rangle ), 
\end{equation}
%$N$ single measurements, 
and $\sigma_{W1}$ and $\sigma_{W2}$ are unbiased variance for $W1$ and $W2$ bands given as, 
\begin{gather}
    \sigma^2_{W1} = \frac{1}{N-1} \sum_i^N (m_{W1,i} - \langle m_{W1}\rangle )^2, \\ 
    \sigma^2_{W2} = \frac{1}{N-1} \sum_i^N (m_{W2,i} - \langle m_{W2}\rangle )^2,
\end{gather}
{respectively, }
where $N$ is the number of epochs used to calculate the $r$, $\sigma^2_{W1}$, and $\sigma^2_{W2}$ value; and $\langle m_{W1} \rangle$ ($\langle m_{W2} \rangle$) is the mean $W1$($W2$) magnitude over all epochs.
%of each sample for each epoch. 

We set the criteria for detecting MIR {variability} as
\begin{equation}
    r>0.75, \ \ \sigma_{W1 }> 0.010,  \ {\rm and} \ \sigma_{W2 }> 0.034. 
\end{equation}

The $r>0.75$ {criterion} is based on basic {statistical} {considerations}. 
We calculate studentized Pearson's $r$ value ($t$) from Pearson $r$ value for each ULIRG with, 
\begin{equation}
    t = r \sqrt{\frac{N-2}{1-r^2}}.
\end{equation}
This $t$ value {follows} $t$-distribution with a degree of freedom of $N - 2$ under the null hypothesis. The $r>0.75$ *{criterion} corresponds to the possibility of < 0.003 \% when $N=22$, which is small enough to {eliminate} contamination from {intrinsically} non-variable sources for our sample size $(\sim 100$). 
{The latter criteria ($\sigma_{W1 }> 0.010$ and $\sigma_{W2 }> 0.034$) {are} {adopted to account for the systematic uncertainties of the $W1$ and $W2$ band photometry (0.010 and 0.034 mag, respectively), as reported in NEOWISE Explanatory Supplement\footnote{https://wise2.ipac.caltech.edu/docs/release/neowise/expsup/index.html (accessed on August 28, 2025)}. }}

{For variable sources, we compare their MIR luminosities with those of supernovae, and adopt a threshold of $\Delta L_{W2} = 10^{43}$ erg s$^{-1}$. 
We consider that sources brighter than this threshold are unlikely to originate from supernova activity since even the most luminous supernova observed in the MIR, SN 2010jl, have peak luminosity of $L_{\rm [3.6\ \mu m]}$ and $L_{\rm [4.5\ \mu m]} \sim 10^{42.3}$ erg s$^{-1}$ \citep{2019ApJS..241...38S}.}

\section{Results}\label{sec:5}
We apply the variability {analysis} explained in Section 4 to our sample. 
We remove 1 object with missing ALLWISE counterpart. 
With the coordinate of the AllWISE source, we obtain NEOWISE photometry. 
In this step, we remove 4 sources without sufficient NEOWISE data. 
The $W2<14.5$ {criterion} further removes 4 ULIRGs. 
We then {calculate} the $\sigma_{W1}$, $\sigma_{W2}$, and $r$ value for 155 ULIRGs.
{Figure 2 shows the overall distribution of $r$, and figure~3 shows the  distribution of $r$ categorized by optical types summarized in \cite{2010MNRAS.405.2505N}.
{The distribution of $r$ shows a clear dependence on optical type. H {\sc ii} galaxies and LINERs are mostly found at lower values ($r \lesssim 0.65$), whereas Seyfert galaxies tend to have higher $r$. 
Seyfert 2s make up the majority of sources with $r > 0.9$, and 9 out of 11 Seyfert 1s also exceed this threshold. 
Although LINERs and H {\sc ii} galaxies generally show weaker correlations, some objects display large variability.
In addition, many ULIRGs fall in the range $r = 0.65$–0.75, and part of ULIRGs that are classified as non-variable may still harbor low-level variability below our detection criteria.}
}

We select the variable sources based on the variable criteria and identified 30 variable ULIRGs. 
{We show light curves of the variable ULIRGs with $\Delta L_{W2} > 10^{43}$ erg s$^{-1}$ in figure 4 and figure 5. } The light curves of variable ULIRGs show diverse behaviors.
We summarize the variability properties of the ULIRGs in table 1. 
{For each variable source}, we calculate $\Delta L_{W2}$, which is listed in table 1 and plotted against the $W1-W2$ color in figure 6. 
We overplot AGN criteria of \cite{2012ApJ...753...30S} in figure 6. 
The $\Delta L_{W2}$ values range from $\Delta L_{W2}$ = $7\times10^{42}$  to $5\times 10^{44}$ erg s$^{-1}$. The vertical line in figure 6 {represents}  $\Delta L_{W2} = 10^{43}$ erg s$^{-1}$, whose value is larger than 
the most luminous supernovae observed in the MIR ($L_{\rm [3.6\ \mu m]}$ and $L_{\rm [4.5\ \mu m]} < 10^{42.3}$ erg s$^{-1}$; \cite{2019ApJS..241...38S}).  
The light curves of the variable ULIRGs with $\Delta L_{\rm{W2}}$ below $10^{43}$ erg s$^{-1}$ are shown in {figure} 7.

\section{Discussion}\label{sec:6}

{The distribution of $r$ differs significantly by optical type: H {\sc ii} galaxies and LINERs are mostly found at lower values ($r \lesssim 0.65$), whereas Seyfert galaxies tend to have higher $r$. This {suggests} that the variability is associated with AGN activity rather than stellar processes, although the influence of sample bias is unknown. We further discuss the origin of the variability in this section. }

\subsection{{Comparison with the Most Luminous {Supernovae} Observed in MIR}}

%\subsubsection{}
{As shown in figure~6, 25 ULIRGs exhibit variability and satisfy $\Delta L_{W2} > 10^{43}$ erg s$^{-1}$. The variability of these 25 ULIRGs cannot be explained by previously known supernovae observed in the MIR. This suggests that variable ULIRGs with $\Delta L_{W2} > 10^{43}$ erg s$^{-1}$ {are not powered by stellar activity,} but by intrinsic AGN luminosity variations or by TDEs. } {We do not attempt to distinguish between AGN and TDE variability in this work.}

\subsection{AGN Signatures}

We examine the observational evidence for AGN activity in the variable ULIRGs with $\Delta L_{\rm W2} > 10^{43}$ erg s$^{-1}$ and summarize the results in table 1. 

Optical classifications from \cite{2010MNRAS.405.2505N} identify 14 out of 25 variable ULIRGs as Seyfert~1 or Seyfert~2 galaxies. 
Among the remaining {11} non-Seyfert galaxies, IRAS 20551-4250, IRAS 12514+1027, and IRAS 09320+6134 (UGC 05101) are identified as AGNs by X-ray observations (\cite{2003MNRAS.343.1181F, 2003MNRAS.338L..19W,2017ApJ...835..179O}). 
{While 8 sources do not show optical or X-ray AGN signatures, all 25 variable ULIRGs except IRAS 01388-4618 satisfy AGN MIR color selection criterion of $W1-W2 > 0.8$ (\cite{2012ApJ...753...30S}). The $W1-W2$ values are summarized in table~1. 
For IRAS 01388-4618, {\it Spitzer} spectroscopy confirms AGN contributions in MIR spectra  (\cite{2010MNRAS.405.2505N}). 
All variable ULIRGs with $\Delta L_{\rm W2} > 10^{43}$ erg s$^{-1}$ fulfill at least one AGN criterion.
In combination with the fact that these ULIRGs have $\Delta L_{\rm W2}$ values that exceed the luminosity of the brightest known supernovae by a factor of $\sim$5. 
The variable ULIRGs with $\Delta L_{\rm W2} > 10^{43}$ erg s$^{-1}$ most likely host AGNs:  even if the observed MIR variability originate from TDEs, the black holes involved are likely active.}

{We cross-matched the ULIRG sample with the eROSITA DR1 variability catalog
(\cite{2025A&A...700A..61B}). 
For all 163 ULIRGs with AllWISE counterparts, no association with an eROSITA DR1 variable source was found within 2$\arcmin$.}

\subsection{Properties}

\subsubsection{Sizes}

{
If the MIR light curve shows a rise and a fall over a timescale $\Delta t$, the radius of the variable {MIR-emitting region $R$ is constrained by 
\begin{equation}
    R \leq c \ \Delta t/2,
\end{equation}
where $c$ is the speed of light.
The $R=c \ \Delta t/2$} corresponds to the case where the driving UV/optical variation is an impulsive (delta-function) flare. If the UV/optical driver varies more slowly, the observed MIR light curve is broadened by light-travel effects, and the true size of the emitting region must therefore be smaller than this upper limit.
For ULIRGs with clear up-and-down light curves over a few years (e.g., IRAS 01572+0009, IRAS 07246+6125, IRAS 08559+1053), the variability timescale implies a compact emitting region with $R \lesssim \ \mathrm{a \ few\ pc}$, which is consistent with the typical size of the hot dust distribution in the inner regions of AGN tori (e.g., \cite{2006ApJ...639...46S,2014ApJ...788..159K,2014ApJ...788...48S}).
We note that variability allows us to constrain only the size of the variable component, while the size of the non-variable component remains unconstrained. Indeed,  
\cite{2024A&A...690A..95A} spatially resolved the extended component of the MIR emission, {from which MIR variability is not expected.}}

{
Some ULIRGs show only a monotonic increase or decrease within $\sim$10 years, in which case no upper limit can be set, following the argument above.
However, it is possible that the inner dust-emitting region also follows a gradual decline over several years, as observed in TDEs or CL AGNs (e.g., \cite{2018ApJ...864...27S}).
{Indeed}, the size of the MIR-emitting region in TDEs lacking optical counterpart is often estimated under the assumption that the TDE light curve follows a $t^{-5/3}$ decline (e.g., \cite{2024ApJ...961..211M}).
Therefore, comparison with optical light curves is important for constraining the size of the variable MIR-emitting region. Such comparisons have already been carried out for several ULIRGs (e.g., \cite{2017NatAs...1E..61T,2016ApJ...828L..14J,2017ApJ...841L...8D,2024A&A...692A.262S}).}

\subsubsection{Rates}
{
We find 30 variable ULIRGs out of 155 ULIRGs, meaning that the incidence rate of variability for this study is around $\sim 20 \ \% $. 
This value is an order of magnitude larger compared to incidence rate of variability found in dwarf galaxies ($\lesssim  1 \% $; \cite{2023ApJ...945..157H,2020ApJ...900...56S,2022ApJ...936..104W}).}

{We also note that our criteria may have excluded objects with small variability amplitudes.
In {figure} 8, we plot $\Delta W2$ versus $\Delta L_{W2}$ for variable sources and non-variable that satisfy only the correlation criterion ($r>0.75$). Variable ULIRGs are concentrated at $\Delta W2 \gtrsim 0.15$, whereas the objects with $r>0.75$ but ULIRGs not satisfying the $\sigma_{W1}$ and $\sigma_{W2}$ thresholds typically show $\Delta W2 \lesssim 0.1$.
Figure~3 already indicates that $r$ value is correlated with AGN activity. However, only three Seyfert 1 galaxies were actually selected as variable under our criteria, despite most Seyfert~1 galaxies showing $r>0.75$. This suggest that our conservative criteria to remove systematics may have led to missing some sources that are actually variable.
For reference, 22 ULIRGs {that show $r>0.75$ but do not satisfy our criteria} are summarized in table~2.
A more careful removal of systematics may increase the incidence rate up to {$\sim 30 \%\ ((30+22)/155)$. }
Even after including the 22 sources, the detection rate of variability in AGNs is larger than in our sample.
%although the AGN studies are based on a shorter time baseline 
(e.g., \cite{2016ApJ...826..118K}).
This suggests that ULIRGs host AGNs that cannot always be detected even in the MIR, or that some ULIRGs lack an AGN or exhibit only low-amplitude variability because their MIR emission is dominated by starburst activity.
{A systematic comparison of MIR variability in ULIRGs with that in normal galaxies and AGNs will  be important.
}}

{
In addition, using AllWISE data \citep{2010AJ....140.1868W}, unTimely data \citep{2023AJ....165...36M}, future observations with SPHEREx \citep{2020SPIE11443E..0IC} would help to confirm additional variability in other ULIRGs and increase the number of MIR variable ULIRGs. 
}

\subsection{Notes for Individual Objects}

\subsubsection{IRAS 05189-2524 and {IRAS 01003-2238}}

{Variable ULIRGs include previously reported TDE candidates in IRAS 01003-2238 {(IRAS~F01004-2237; \cite{2017ApJ...841L...8D,2024A&A...692A.262S}) and IRAS~05189-2524 \citep{2022A&A...664A.158R}.}
Both of these candidates are successfully detected with our method, demonstrating that our approach is also effective for identifying TDE candidates.  }

{
IRAS~05189-2524 also shows a luminosity increase in the last few epochs apart from the initial burst, which may correspond to another transient event.  
In {figure~8} we compare the variability properties of ULIRGs. No ULIRGs occupy the same region of parameter space as {IRAS 01003-2238} (which has $\Delta W2 > 0.4$~mag). In contrast, several variable ULIRGs overlap with the parameter space of {IRAS~05189-2524}. Such systems may already be detectable with optical time-domain surveys (ZTF, ATLAS, Pan-STARRS). However, many of these analogs are likely fainter than {IRAS~05189-2524} and may fall below the detection limits of current surveys ({figure} 8). Deeper monitoring with the Rubin Observatory \citep{2019ApJ...873..111I} should uncover part of such optical variability of ULIRG.
}

\subsubsection{IRAS 00456-2904 and  IRAS 17028+5817}
{Five objects in our sample have $\Delta L_{\rm{W2}}$ below $10^{43}$~erg~s$^{-1}$.
Among them, two galaxies, IRAS~00456–2904 and IRAS~17028+5817, lack optical, X-ray, MIR color, or {MIR-spectroscopic} AGN signatures. 
{For IRAS 00456–2904, ALMA observations reveal an elevated HCN-to-HCO$^+$ flux ratio, suggesting the presence of an optically elusive AGN (\cite{2023ApJ...954..148I}).
For IRAS 17028+5817, we do not find AGN signatures reported in the literature.
Notably, this source exhibits two apparent transient events whose light curves superficially resemble those of repeating partial TDEs (\cite{2025ApJ...979...40L}),
with $\sigma_{W1}$ smaller than  $\sigma_{W2}$ ($\sigma_{W2}/\sigma_{W1} = 3.7$). 
This may indicate that the infrared emission itself is attenuated by dust. 
The properties of IRAS~17028+5817 will be discussed in a forthcoming paper. }}

\subsubsection{{IRAS~01003-2238 and IRAS~01572+0009}}
{IRAS~01003-2238 and IRAS~01572+0009 (Mrk 1014) show repeated brightening and fading on timescales of a few years. Both ULIRGs have been suggested to host dual AGNs based on SED fitting by \cite{2022MNRAS.512.5183E}, which could be responsible for enhanced nuclear activity.}

\begin{figure}
 \begin{center}
  \includegraphics[width=7cm]{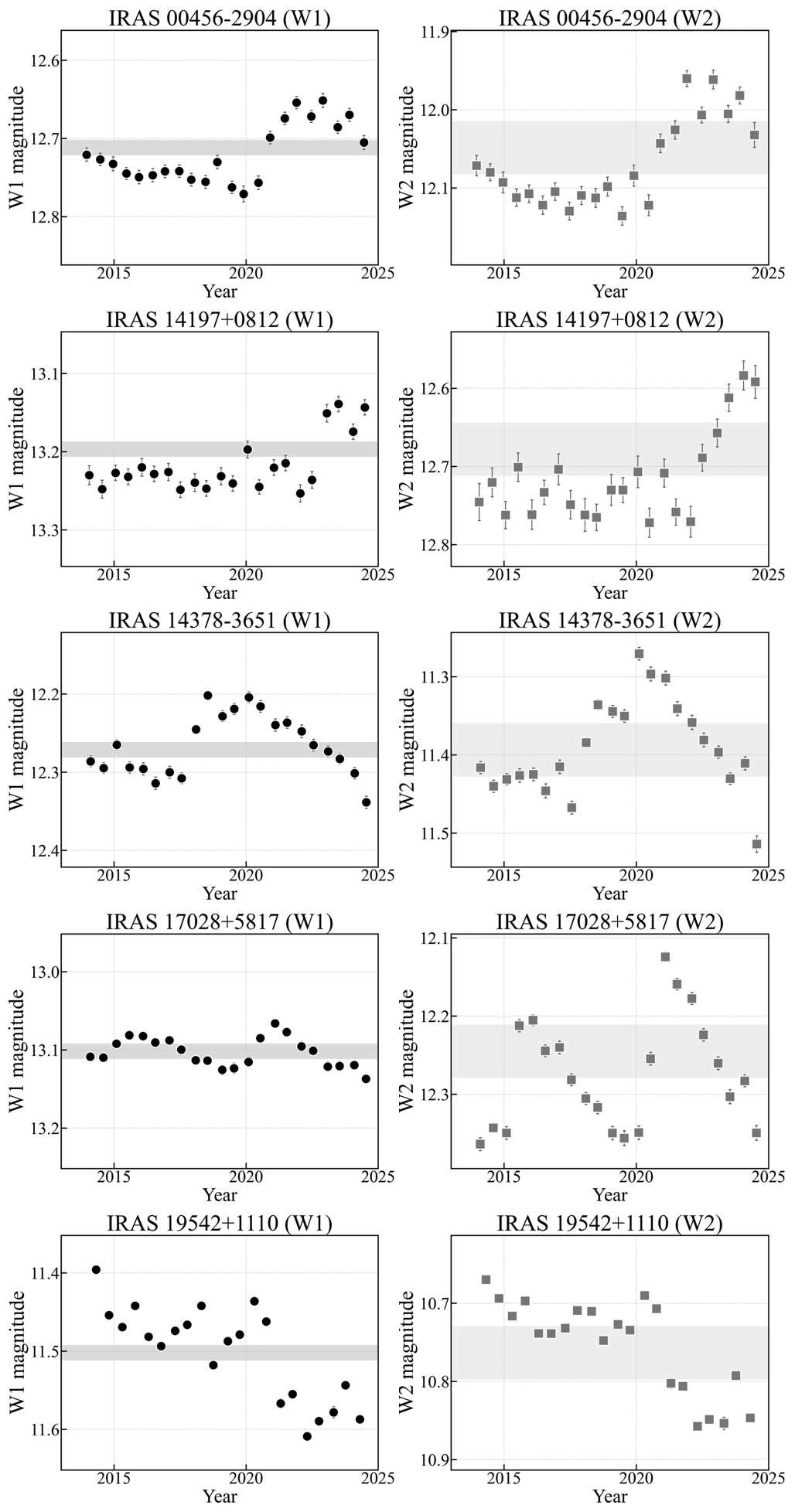} 
 \end{center}
\caption{
{Binned $W1$ (black circles) and $W2$ (gray squares) band light curves of variable ULIRGs with $L_{\rm W2}<10^{43}$ erg s$^{-1}$, which are not the main focus of this study. The gray horizontal band indicates twice the systematic uncertainty for each band. 
{Alt text: Binned W1 and W2 light curves for five galaxies with W2 peak luminosity below ten to the forty three erg per second. Each galaxy is displayed in two panels with magnitude on the vertical axis and time on the horizontal axis. The light curves show variability, but these objects are presented only as reference.}
}
} 
\label{fig:sample}
\end{figure}

\subsection{Precise Physical Origins of Variability}

{The observed MIR variability in ULIRGs can arise from either thermal or non-thermal emission. Thermal emission originates from the reprocessing of UV/optical radiation by surrounding dust, while non-thermal emission corresponds to synchrotron radiation from relativistic jets, such as in blazars. 
Blazers often show variability on timescale of hours to days (e.g., \cite{2012ApJ...759L..31J,1996ASPC..110..391U}), 
{which is shorter than that of thermal emission, as the latter is smoothed due to the relatively larger geometry of the thermal IR–emitting region.}
A few ULIRGs in our sample (e.g., IRAS 11223-1244) show hectic up and down variability. This may arise from non-thermal emission from the jets, however, dust which reprocess UV photon may be distributed in compact region {($\lesssim 0.1$ pc $\sim  1/2 $ ly)}. 
{High-cadence} observation or {simultaneous} observation in radio wavelength is {effective} to {disentangle} the origin of such ULIRGs.
For the other sources, the MIR variability show relatively smooth transition, and is more consistent with thermal reprocessing.
Thus, we mainly focus on thermal emission origin hereafter.}

{The smooth MIR variability originate from the infrared echo of the incident UV/optical emission or obscuration of MIR emitting region, while due to the large MIR luminosities and spatial scales of these systems, variability due to obscuration is less plausible. 
Variations likely result from intrinsic changes in the accretion power. 
Several physical mechanisms can produce intrinsic variability.
These include stochastic fluctuations in AGN accretion flows, TDEs, and transitions in accretion disk states {as} discussed in CLAGNs.
Stochastic fluctuations in AGNs show hectic up and down light curves, rather than {a} single flaring event seen in TDEs and CLAGNs. 
For TDEs and CLAGNs, TDEs typically exhibit short-duration, high-amplitude bursts, whereas CLAGNs show smooth, long-lasting optical variations that {are} driven by substantial changes in the accretion rate or disk structure
(\cite{2016MNRAS.457..389M,2018ApJ...864...27S}). }{There are several objects with detected intrinsic UV/optical flare varying on short timescales (IRAS 01003-2238 and IRAS 05189-2524). For these objects, the origin of MIR variability is TDE candidates \citep{2017ApJ...841L...8D,2022A&A...664A.158R}.}

{{For variable ULIRGs with optical variability or flaring events not reported yet}, there is currently no firm evidence for the underlying origin of the MIR variability. Hereafter, we discuss {only the } MIR light curves. }

{
For several objects, we investigate in further detail whether the observed variability can be explained by a single flare event or requires multiple flares.
We focus on ULIRGs with unique features observed in their light curves: IRAS 01003-2238, IRAS 01572+0009, and IRAS 12071-0444. 
These ULIRGs show slow rising or double-peaked MIR light curves. 
Theoretically, these can be reproduced by geometrically thick torus models {\citep{2017ApJ...841L...8D}}, where the extended dust distribution causes temporally broadened echoes.
Although simple torus models have been successful in reproducing MIR flares of either gradual increase or slow decline (e.g., \cite{2022ApJS..258...21W,2024ApJ...961..211M}), 
they have difficulty reproducing both the gradual rise and subsequent slow decline observed in some IR echoes, {as} seen in IRAS 01003-2238, IRAS 12071-0444, and IRAS 10572+0009 (e.g., \cite{2017ApJ...841L...8D}). }
{ 
{An} alternative idea for explaing these features are off-center UV/optical flares, for instance from one component of a binary SMBH system. 
The resulting asymmetric illumination could also produce a gradual rise and decline. }
{It is possible that binary SMBHs reside in ULIRGs, most of which are interacting or merging galaxies.
AGN activity is thought to be triggered by merger events, as suggested by both theoretical and observational studies (e.g., \cite{2008ApJS..175..356H,2014MNRAS.441.1297S}).
During the merging process, a binary SMBH can remain at a separation of $\sim$kpc for an extended period before its orbit gradually decays to the pc scale through dynamical interactions. The binary can persist for a long time at this stage before gravitational-wave emission becomes efficient at milliparsec separations (e.g., \cite{2015MNRAS.447.2123C}).
Simulations suggest that AGNs with higher luminosities during mergers tend to have smaller SMBH separations (\cite{2012ApJ...748L...7V}). Combined with the large variability amplitudes observed in our ULIRG sample, these {imply} that some of our variable ULIRGs may host binary SMBHs.}
{For binary SMBHs, periodic increase of accretion rate is predicted from simulation \citep{2008ApJ...672...83M}. These may be responsible in part for the high {occurence} rate of MIR variability in ULIRGs. 
}

{
To explore this idea, we construct a toy model of a dust echo considering only geometrical delay from a geometrically thick, optically thin torus with a rectangular cross section, as illustrated in Figure~9. 
The model has a characteristic radius $R_0$, inner and outer radii $R_{\rm in}$ and $R_{\rm out}$, a vertical half-height $H$, and an inclination angle $i$, where $R_0 = (R_{\rm out}+R_{\rm in})/2$.
The illuminating source is offset from the torus center by a distance $r_0$ in the direction of the azimuthal angle $\phi_0$.
}

{We first calculate the geometrical delay for each dust element at position $\mathbf{r}$ as
\begin{equation}
\tau(\mathbf{r}) =
\frac{
|\mathbf{r}-\mathbf{r}_{\rm s}|
-
(\mathbf{r}-\mathbf{r}_{\rm s})\cdot \mathbf{n}_{\rm obs}
}{c},
\end{equation}
where $\mathbf{r}_{\rm s}$ is the position of the illuminating source and $\mathbf{n}_{\rm obs}$ is the unit vector pointing toward the observer.
We weight each dust element by the incident flux from the source, $w(\mathbf{r}) \propto |\mathbf{r}-\mathbf{r}_{\rm s}|^{-2}$.
The transfer function is then constructed as $\Psi(\tau) \propto \sum_j w_j \delta(\tau-\tau_j)$ and is normalized to satisfy $\int \Psi(\tau)\,d\tau = 1$.
}

{
We assume that the intrinsic UV/optical flare has a Gaussian profile, $G(t;\Delta t_{\rm fl})$.
Here, $\Delta t_{\rm fl}$ is the full width at half maximum of the Gaussian and represents the duration of the flare.
The observed MIR light curve is modeled as the dust echo produced by convolving this flare profile with the transfer function $\Psi$.
The model is written as
\[
F_{\rm model}(t)
=
A \int \Psi(\tau)\,G(t-T_0-\tau;\Delta t_{\rm fl})\,d\tau + B,
\]
where $T_0$ is the time offset of the flare relative to the first observed epoch, $A$ is a normalization factor, and $B$ represents a constant MIR component.
For a given set of torus parameters, $A$, $B$, and $T_0$ are determined by minimizing $\chi^2$ between the model and the observed light curve.
}

{We manually explore the torus parameters to find models that reproduce the broad rise and decay features of the observed MIR variability.
As a first step, we perform this search assuming that the illuminating source is located at the torus center.
The resulting model light curves are shown in Figure~10, and the corresponding model parameters are summarized in table~\ref{tab:params}.
These models successfully reproduce the overall shapes of the light curves for the three flares.
For {IRAS 12071-0444}, however, the centered-source model does not fully reproduce the double-peaked structure of the observed light curve.
We therefore additionally construct an offset-source model for {IRAS 12071-0444}, shown by the black line in Figure~10.
This model better reproduces the double-peaked emission feature.
We also summarize corresponding model parameters in table~\ref{tab:params}.
}

{The offset-source model is not unique, because similar MIR light curves could also be produced by assuming a more complex UV/optical input flare rather than a single Gaussian flare.
Nevertheless, the model illustrates that even a single UV/optical flare can reproduce the double-peaked MIR feature when the illuminating source is displaced from the torus center.}

{
If we assume that the inner radius of the torus corresponds to the dust sublimation radius, we can estimate the UV luminosity associated with this radius.
We denote this luminosity as $L_{\rm UV,sub}$.
We use the dust sublimation radius relation given by \citet{2007A&A...476..713K}, originally based on \citet{1987ApJ...320..537B},
\[
R_{\rm sub}
=
1.3
\left(\frac{L_{\rm UV}}{10^{46}\ {\rm erg\ s^{-1}}}\right)^{1/2}
\left(\frac{T_{\rm sub}}{1500\ {\rm K}}\right)^{-2.8}
\left(\frac{a}{0.05\ \mu{\rm m}}\right)^{-1/2}
{\rm pc}.
\]
Here, $T_{\rm sub}$ is the dust sublimation temperature and $a$ is the dust grain size.
We adopt $T_{\rm sub}=1500\,{\rm K}$ and $a=0.05\,\mu{\rm m}$.
We calculate $L_{\rm UV,sub}$ by setting $R_{\rm sub}=R_{\rm in}$.
For comparison, we also list the total IR luminosities $L_{\rm IR}$ of the ULIRGs from \citet{2010MNRAS.405.2505N} and the ratios $L_{\rm IR}/L_{\rm UV,sub}$ in table~\ref{tab:params}.
The resulting ratios are $L_{\rm IR}/L_{\rm UV,sub}=0.1$--$0.4$.
Because the listed $L_{\rm IR}$ values are total galaxy IR luminosities and may include contributions from star formation, we regard $L_{\rm IR}/L_{\rm UV,sub}$ as an upper limit on the fraction of the UV luminosity reprocessed into IR emission.
These values indicate that the inferred UV luminosities are energetically sufficient to power the observed IR luminosities, if the luminosity is sustained over time.
However, because the inferred $R_{\rm in}$ may reflect a dust sublimation radius that temporarily expanded during the flare rather than the steady-state inner radius of the torus, this comparison should be regarded as an order-of-magnitude energetic check.
}

\section{Summary}\label{sec:6}

{We analyzed 11 years of NEOWISE photometry for 164 ULIRGs from \cite{2010MNRAS.405.2505N} to investigate MIR variability as a probe of buried AGNs.
We detected significant variability in 30 ULIRGs, among which 25 show amplitudes exceeding $\Delta L_{W2}=10^{43}$ erg s$^{-1}$, brighter than the most luminous SNe known in the MIR.
These include previously reported TDE candidates and hard X-ray detected AGNs.
Because the variability is observed on timescales of a few years, the MIR-emitting region is suggested to be compact, with sizes of order $\lesssim \mathrm{a \ few\ pc}$ for ULIRGs with both increase and decrease of the luminosity. 
The large amplitudes cannot be explained by SNe, and are most likely due to intrinsic AGN luminosity changes or TDEs occurring in AGN-hosting systems. 
All variable ULIRGs with $\Delta L_{W2}>10^{43}$ erg s$^{-1}$ show at least one independent AGN signature, supporting this interpretation.
This work demonstrates that MIR variability can define a new field in uncovering buried nuclear activity in ULIRGs.
}

%%%%%%%%%%%%%%%%%%%%%%%%%%%%%%%%%%%%%%% 

% See the instruction below for "Alt text"
% https://academic.oup.com/pasj/pages/General_Instructions#Figures%20and%20Illustrations

\begin{table*}
  \tbl{Variable Sources.\footnotemark[$*$] }{%
 \begin{tabular}{l c c c c c c c c c c c c}
\hline
{\it IRAS} name & $z$ & $r$ & $\sigma_{W1}$ & $\sigma_{W2}$ & Abol (\%)  & $W2$ & $W1\! -\! W2$ &$\log_{10}\Delta L_{W2}\ (\mathrm{erg\ s^{-1}})$&
 Type & AGN sign(s)? & Notes & \\
(1)& (2)& (3) & (4) & (5) & (6) & (7) & (8) & (9) & (10) & (11) & (12) \\
\hline
00275-0044 & 0.242   & 0.763 & 0.039 & 0.079 & 3.2 & 12.81 &1.05 & 43.52 &       & Y(a,b)&\\
01003-2238 & 0.118  & 0.952 & 0.207 & 0.140 & 50.  & 11.26 &1.63 & 44.08 & H {\sc ii}  & Y(a,b) & TDE candidates (f,g) \\
01388-4618 & 0.090 & 0.954 & 0.028 & 0.067 & 1.6   & 11.43 &0.61 & 43.12 & H {\sc ii}  & Y(a)&   \\
01572+0009 & 0.163 & 0.974 & 0.143 & 0.108 & 27.   & 9.79  &1.00 & 44.45 & Sy 1  & Y(a,b)& Mrk 1014\\
05189-2524 & 0.043 & 0.979 & 0.055 & 0.036 & 30.   & 7.52  &1.13 & 43.78 & Sy 2  & Y(a,b)& TDE candidate (h) \\
06361-6217 & 0.160  & 0.986 & 0.071 & 0.042 & 18.  & 10.86 &1.75 & 43.70 &       & Y(a,b)&  \\
07246+6125 & 0.137 & 0.932 & 0.064 & 0.050 & 19.   & 12.06 &1.15 & 43.21 & Sy 2  & Y(a,b)& \\
07572+0533 & 0.190 & 0.784 & 0.099 & 0.100 & 25.   & 12.33 &1.11 & 43.63 & LINER & Y(a,b)&  \\
08559+1053 & 0.148 & 0.873 & 0.048 & 0.038 & 7.6   & 10.49 &1.06 & 43.85 & Sy 2  & Y(a,b)&  \\
09320+6134 & 0.039 & 0.979 & 0.042 & 0.049 & 14.   & 8.34  &1.70 & 43.42 & LINER & Y(a,b)& UGC 05101 (h), X-ray AGN (i)\\
11223-1244 & 0.199 & 0.884 & 0.044 & 0.038 & 5.6   & 12.06 &1.09 & 43.36 & Sy 2  & Y(a,b)& \\
12071-0444 & 0.128 & 0.988 & 0.100 & 0.085 & 41.   & 10.71 &1.34 & 43.84 & Sy 2  & Y(a,b)& \\
12514+1027 & 0.300 & 0.975 & 0.034 & 0.035 & 96.   & 9.73  &1.30 & 44.63 &       & Y(a,b)&  X-ray AGN (j)\\
12540+5708 & 0.042 & 0.908 & 0.117 & 0.075 & 34.   & 6.30  &1.10 & 44.60 & Sy 1  & Y(a,b)& Mrk 231 \\
15130-1958 & 0.109 & 0.992 & 0.109 & 0.088 & 30.   & 10.01 &1.23 & 43.84 & Sy 2  & Y(a,b)& \\
15176+5216 & 0.139 & 0.960 & 0.071 & 0.041 & 37.   & 9.85  &1.19 & 43.93 & Sy 2  & Y(a,b)& \\
15462-0450 & 0.100 & 0.957 & 0.091 & 0.061 & 26.   & 9.67  &1.17 & 43.87 & Sy 1  & Y(a,b)& \\
16155+0146 & 0.132 & 0.988 & 0.062 & 0.070 & 40.   & 10.71 &2.39 & 43.63 & Sy 2  & Y(a,b)& \\
16334+4630 & 0.191 & 0.932 & 0.041 & 0.043 & 1.0   & 12.19 &1.11 & 43.35 & LINER & Y(a,b)&  \\
16541+5301 & 0.194 & 0.938 & 0.064 & 0.065 & 6.0   & 12.59 &0.84 & 43.40 & Sy 2  & Y(a,b)& \\
17044+6720 & 0.135 & 0.983 & 0.043 & 0.044 & 27.   & 10.44 &1.53 & 43.64 & LINER & Y(a,b,c)& \\
19254-7245 & 0.062 & 0.954 & 0.044 & 0.049 & 24.   & 8.80  &1.70 & 43.65 & Sy 2  & Y(a,b)& Superantenna\\
20087-0308 & 0.106 & 0.942 & 0.031 & 0.057 & 3.1   & 10.78 &1.23 & 43.40 & LINER & Y(a,b)&  \\
20551-4250 & 0.043 & 0.952 & 0.033 & 0.109 & 26.   & 9.83  &1.44 & 43.25 & LINER & Y(a,b)& X-ray AGN (k)\\
23498+2423 & 0.212 & 0.957 & 0.080 & 0.072 & 26.   & 10.36 &1.15 & 44.32 & Sy 2  & Y(a,b)& \\
\hline
00456-2904 & 0.110 &  0.951 & 0.037 & 0.056 & ---  & 12.05 &0.68 & 42.93 & H {\sc ii}  & Y(d)& \\
14197+0812 & 0.131 &  0.843 & 0.035 & 0.057 & 11.  & 12.71 &0.49 & 42.87 &       & Y(a,e)&\\
14378-3651 & 0.068 &  0.920 & 0.039 & 0.060 & 1.0  & 11.38 &0.87 & 42.89 & Sy 2  & Y(a,b)&\\
17028+5817 & 0.106 &  0.779 & 0.019 & 0.071 & ---  & 12.32 &0.78 & 42.95 & LINER & N&\\
19542+1110 & 0.065 &  0.981 & 0.060 & 0.060 & 3.8  & 10.76 &0.76 & 42.99 &       & Y(a)& \\
\hline
\end{tabular}
}\label{tab:first}
\begin{tabnote}
\footnotemark[$*$] 
{General properties and statistics of our variable ULIRG sample. 
(1) {\it IRAS} name, (2) redshift , (3) Pearson $r$ value,
(4) Standard variation of $W1$ photometric values, (5) Standard variation
of $W2$ photometric values, (6) AGN bolometric contribution (in per cent) taken from \cite{2010MNRAS.405.2505N}, 
(7) $W2$ magnitude taken from AllWISE catalog, (8) {\it WISE} $W1-W2$ color, (9) Difference between max and minimum $W2$ band
luminosity, (10) optical class summarized by \cite{2010MNRAS.405.2505N}, (11) presence of AGN signatures (Y = yes, N = no), and several selected representative references, and (12) notes. }
(a) \cite{2010MNRAS.405.2505N} {(SED fitting to MIR spectroscopic data)} , (b) \cite{2012ApJ...753...30S} {(MIR color)},  (c) \cite{2014ApJ...780..106I} {(MIR color)}, (d) \cite{2023ApJ...954..148I} {(enhanced HCN/CO+ emission line ratio)}, (e) \cite{2006AJ....131.2406I} {(SED fit to MIR spectra)}, (f) \cite{2017ApJ...841L...8D}, (g) \cite{2024A&A...692A.262S}, 
(h) \cite{2022A&A...664A.158R}, (i) \cite{2017ApJ...835..179O}, (j) \cite{2003MNRAS.338L..19W}, (k) \cite{2003MNRAS.343.1181F}.
\end{tabnote}
\end{table*}

\begin{figure*}
 \begin{center}
  \includegraphics[width=15cm]{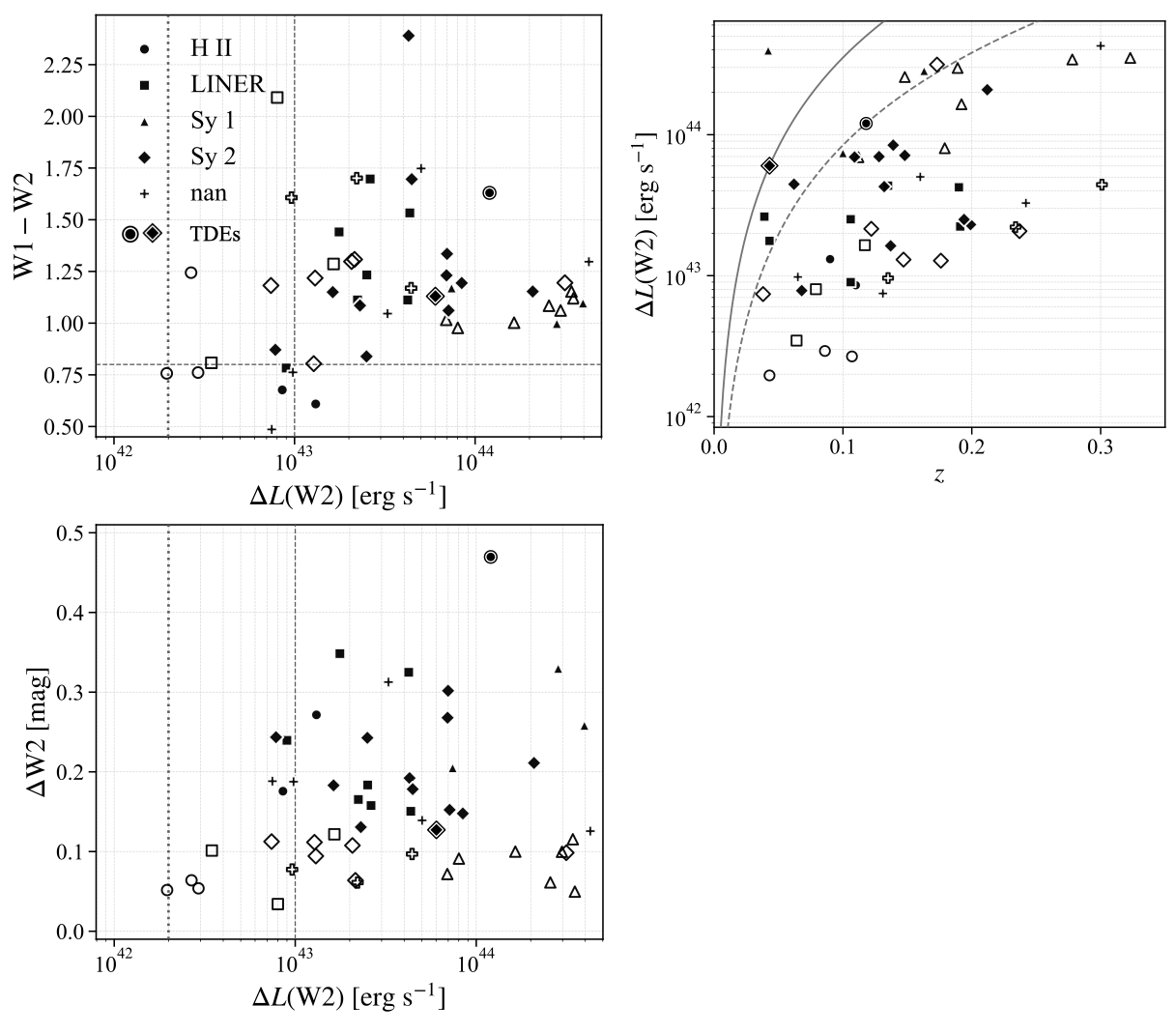} 
 \end{center}
\caption{
{Property of variable ULIRGs. Relation between difference of  color, and $\Delta L_{W2}$. 
 Upper left panel: $W2$ magnitude ($\Delta W2$), $W1-W2$ for different Type. {The vertial dotted line, dashed line, and horizontal dashed line are the same as {figure} 6.}
 Lower panel: $\Delta W2$ as a function of  \ $\Delta {W2}$. {{The vertial dotted and dashed lines are the same as {figure} 6.}}
 Upper right panel: redshift as a function of $\Delta L_{W2}$. 
Filled symbols mark sources classified as variable; open (white) symbols indicate sources with $r > 0.75$ but do not satisfy the variability thresholds.
Optical spectroscopic type is encoded by marker shape. 
Known TDE candidates are highlighted with double circles and double squares.
Gray solid and dashed lines in the upper-right panel show $\Delta L_{W2}$ value  that would appear with the same brightness at different redshifts.
The rest are the same as {figure} 6.}
{Alt text: Figure showing properties of variable ULIRGs and their relations between color and $\Delta L_{W2}$. The figure consists of three panels. The upper left panel plots $\Delta W2$ versus $W1-W2$ color, showing different optical types with distinct marker shapes. Vertical dotted and dashed lines, as well as a horizontal dashed line, are the same as in figure 6. The lower panel shows $\Delta W2$ as a function of $\Delta {W1-W2}$, with vertical dotted and dashed lines also identical to those in figure 6. The upper right panel presents redshift as a function of $\Delta L_{W2}$. Filled symbols indicate variable sources, while open symbols mark sources with $r>0.75$ that do not meet the variability threshold. Known TDE candidates are emphasized with double circles or double squares. Gray solid and dashed lines indicate the $\Delta L_{W2}$ values corresponding to the same apparent brightness at different redshifts. All other notations are consistent with figure 6. The lower panel demonstrates that sources with variability but satisfying $r > 0.75$ generally exhibit small $\Delta W2$ values.
The upper left panel shows that some ULIRGs have properties similar to the known TDE candidates.
The upper right panel, however, indicates that the known TDE candidates are characterized by being both nearby and luminous.}
}
\label{fig:sample}
\end{figure*}

\begin{figure}
 \begin{center}
  \includegraphics[width=7cm]{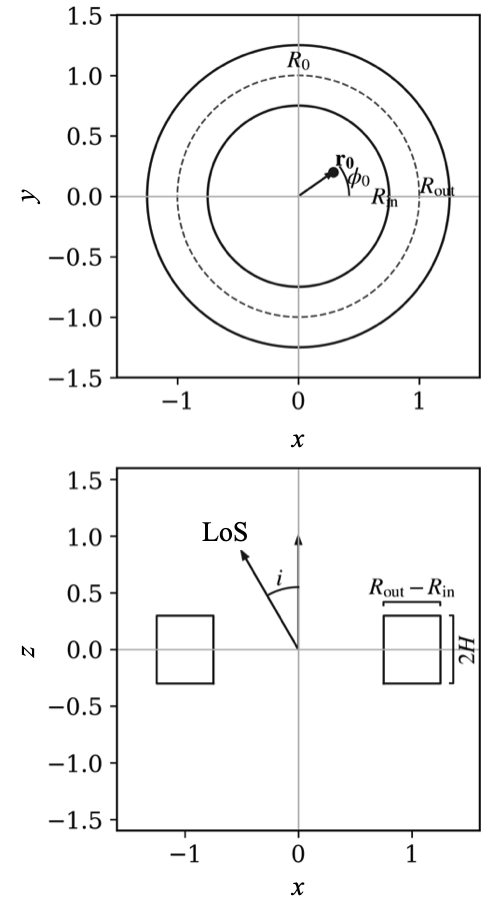} 
 \end{center}
\caption{
{Cartoon of the dust torus geometry. The upper panel shows the torus viewed face-on, while the lower panel shows the edge-on view (slice at y = 0). The axes are shown in units of $R_0$.
Alt text:
Cartoon illustrating the dust torus geometry.
The upper panel shows a face-on view, and the lower panel shows an edge-on cross section at $y = 0$.
}}
\label{fig:sample}
\end{figure}

\begin{figure}
 \begin{center}
  \includegraphics[width=8cm]{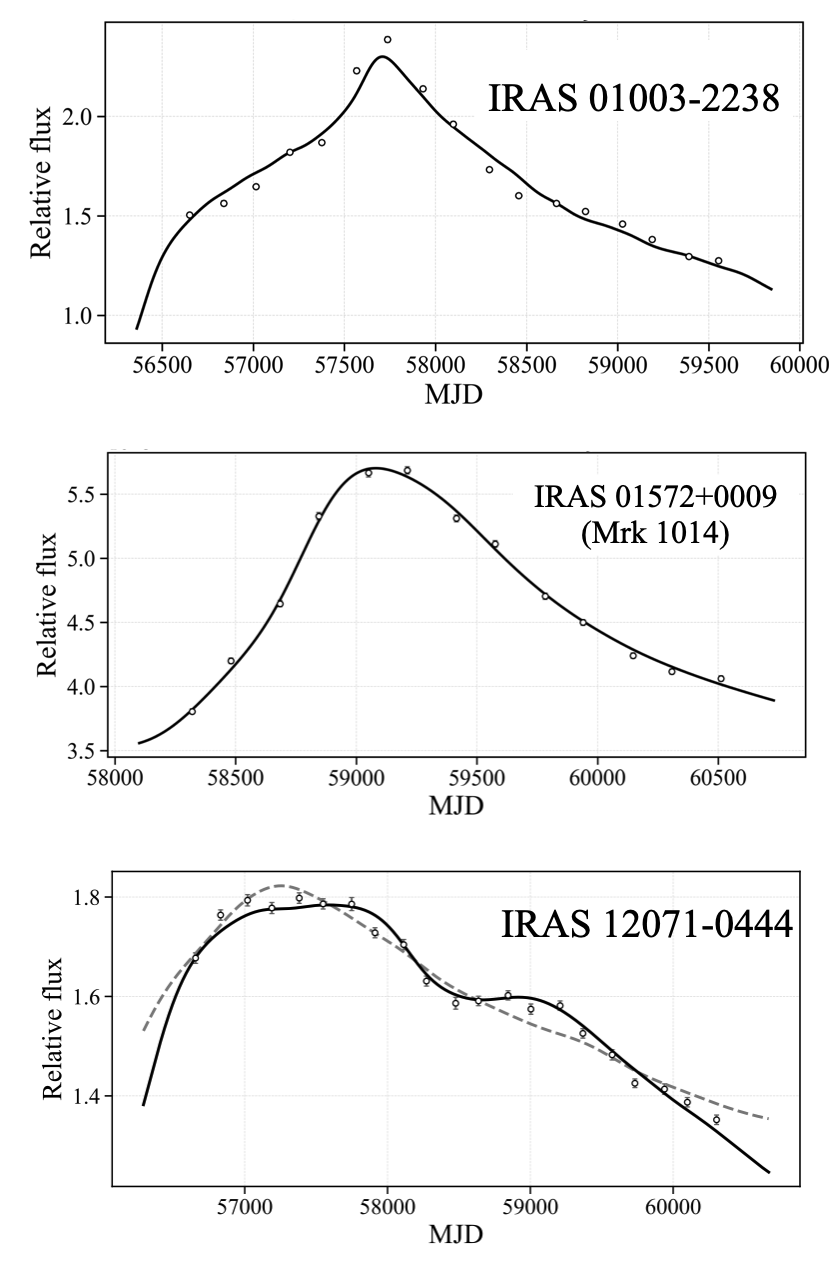} 
 \end{center}
\caption{
{
Examples of theoretical light curves for selected variable ULIRGs.
{For IRAS 01003-2238 and IRAS 01572+0009,} solid lines show dust echo models, and open circles show the observed $W1$-band fluxes with their associated uncertainties, in arbitrary units.
For IRAS 12071-0444, the black solid line shows an offset-source model, while the gray solid {dashed line} shows a centered-source model.
Alt text: Examples of theoretical light curves compared with observed $W1$-band data for selected variable ULIRGs.
Open circles with error bars show the observed data, and solid lines show dust echo models.
For IRAS 12071-0444, the black and gray solid lines show the offset-source and centered-source models, respectively.
}
}

\label{fig:sample}
\end{figure}

%%%%%%%%%%%%%%%%%%%%%%%%%%%%%%%%%%%%%%%

\begin{ack}
{This study is based on the group work carried out at the "Galaxy-IGM Workshop 2023" {held} in Hamamatsu. We thank the organizers for providing this valuable opportunity. The authors thank the Yukawa Institute for Theoretical Physics at Kyoto University. Discussions during the YITP workshop YITP-W-25-08 on "Exploring Extreme Transients" were useful to complete this work.
We thank Takumi S. Tanaka, Kohei Inayoshi, Takashi J. Moriya, {Takeo Minezaki}, {Toni Tuominen}, Hiroya Umeda, Masami Ouchi, Yusei Koyama, Hirofumi Noda, Yuichi Harikane, Tomokazu Kiyota, Shogo Yoshioka, Shohei Aoyama, Nanase Harada, and Shigeo Kimura for helpful discussions and useful suggestions. 
{We thank the anonymous referee for their valuable and constructive comments that have greatly improved the quality of the manuscript.}
{We thank Junya Arita, Ryota Ikeda, Kimihiko Nakajima, Rui Marques-Chaves, Anne Verhamme, Tadashi Nakajima, Kazunori Kohri, Haruka Kusakabe, Roberto Maiolino, Xihan Ji, Kirk Long, Ryotaro Chiba, Kohei Ichikawa, Taiki Kawamuro, Mitsuru Kokubo, Takato Tokuno, Takatoshi Ko, Hiroshi Nagai, Luis C. Ho, Shinobu Ozaki, Chihiro Tokoku, Yuma Sugahara, Yuki Kambara, Yutaro Kuroda, Kentaro Nagamine, Kazuhiro Hada, Kazunori Shimasaku, and Yoshinobu Fudamotom for their kind concern and encouragement. 
We thank Yuki Sheena and Pickels Taeko for helpful guidance on figures and color.
}}
This publication makes use of data products from the Wide-field Infrared Survey Explorer, which is a joint project of the University of California, Los Angeles, and the Jet Propulsion Laboratory/California Institute of Technology, and NEOWISE, which is a project of the Jet Propulsion Laboratory/California Institute of Technology. WISE and NEOWISE are funded by the National Aeronautics and Space Administration.
This research has made use of the NASA/IPAC Infrared Science Archive, which is funded by the National Aeronautics and Space Administration and operated by the California Institute of Technology.
\end{ack}

\section*{Funding}
 S.H is supported by JSPS KAKENHI grant No. 24KJ1159. M.I. is supported by JP21K03632 and JP25K07359. T.K. is supported by JP22K14076 and {JP26K07133}.

\section*{Data availability} 
 The data underlying this article are available on NASA/IPAC Infrared Science Archive (https://irsa.ipac.caltech.edu/frontpage/).

\appendix %%%%%%%%%%%%%%%%%%%%%%%%%%%%%%%%%%%%%%%%%%%%%%%%%%%%%%%%

\begin{table*}
  \tbl{ULIRGs not selected as variable source but with $r>0.75$.\footnotemark[$*$] }{%
 \begin{tabular}{l c c c c c c c c c c c c}
\hline
{\it IRAS} name & $z$ & $r$ & $\sigma_{W1}$ & $\sigma_{W2}$ & Abol (\%)  & $W2$ & $W1\! -\! W2$ &$\log_{10}\Delta L_{W2}\ (\mathrm{erg\ s^{-1}})$&
 Type \\
 \hline
00275-2859 & 0.278 & 0.91 & 0.037 & 0.031 & 78 & 9.77 & 1.15 & 44.53 & Sy 1 \\
03538-6432 & 0.301 & 0.79 & 0.027 & 0.027 & 14 & 11.95 & 1.17 & 43.65 & \\
04103-2838 & 0.117 & 0.95 & 0.037 & 0.034 & 5.4 & 11.04 & 1.29 & 43.22 & LINER \\
04394-3740 & 0.237 & 0.90 & 0.029 & 0.033 & 21 & 12.30 & 1.30 & 43.32 & Sy 2 \\
06035-7102 & 0.079 & 0.87 & 0.012 & 0.010 & 22 & 9.52 & 2.09 & 42.90 & LINER \\
07598+6508 & 0.148 & 0.94 & 0.016 & 0.019 & 85 & 7.84 & 1.08 & 44.41 & Sy 1 \\
10565+2448 & 0.043 & 0.76 & 0.010 & 0.012 & -- & 10.16 & 0.76 & 42.29 & H \sc {ii} \\
11119+3257 & 0.189 & 0.93 & 0.033 & 0.026 & 46 & 8.85 & 1.06 & 44.47 & Sy 1 \\
13342+3932 & 0.179 & 0.90 & 0.032 & 0.022 & 18 & 10.04 & 0.98 & 43.90 & Sy 1 \\
13352+6402 & 0.234 & 0.83 & 0.015 & 0.014 & 12 & 11.62 & 1.70 & 43.34 &  \\
13428+5608 & 0.038 & 0.96 & 0.028 & 0.026 & 7.5 & 9.23 & 1.18 & 42.87 & Sy 2 \\
13451+1232 & 0.122 & 0.79 & 0.013 & 0.019 & 60 & 10.22 & 1.31 & 43.33 & Sy 2 \\
14026+4341 & 0.323 & 0.80 & 0.017 & 0.013 & 92 & 9.28 & 1.12 & 44.54 & Sy 1 \\
17179+5444 & 0.147 & 0.88 & 0.014 & 0.022 & 17 & 11.51 & 1.22 & 43.11 & Sy 2 \\
18443+7433 & 0.135 & 0.91 & 0.017 & 0.020 & 23 & 11.51 & 1.61 & 42.98 & \\
18580+6527 & 0.176 & 0.92 & 0.018 & 0.028 & 1.4 & 12.30 & 0.80 & 43.11 & Sy 2 \\
18588+3517 & 0.107 & 0.75 & 0.018 & 0.018 & 3.5 & 12.11 & 1.24 & 42.43 & H {\sc ii}\\
19297-0406 & 0.086 & 0.85 & 0.014 & 0.015 & 1.0 & 11.28 & 0.76 & 42.47 & H {\sc ii} \\
20037-1547 & 0.192 & 0.98 & 0.035 & 0.032 & 26 & 9.46 & 1.00 & 44.22 & Sy 1 \\
21219-1757 & 0.112 & 0.94 & 0.024 & 0.019 & 84 & 8.93 & 1.02 & 43.84 & Sy 1 \\
23060+0505 & 0.173 & 0.93 & 0.043 & 0.034 & 78 & 8.61 & 1.19 & 44.50 & Sy 2 \\
23365+3604 & 0.064 & 0.83 & 0.018 & 0.024 & 7.2 & 11.16 & 0.81 & 42.54 & LINER \\
\hline
\end{tabular}
}\label{tab:first}
\begin{tabnote}
\footnotemark[$*$] 
{The same as table 1.}
\end{tabnote}
\end{table*}

\begin{table*}
\centering
\caption{
{Geometric parameters of the {representative torus echo models} for the selected variable ULIRGs analyzed in this work, as well as the {UV luminosity required to set the inferred inner dust radius (\(L_{\rm UV,sub}\))} and the luminosity ratio ($L_{\rm IR}/{L_{\rm UV,sub}}$).
The bolometric luminosity is estimated from the inner radius of the torus model, following \cite{2007A&A...476..713K}, and $L_{\rm IR}$ values are taken from \cite{2010MNRAS.405.2505N}.}
}
\label{tab:params}
\begin{tabular}{lrrrrrrrrrr}
\hline
Target & $i$ & $r_0/R_0$ & $\phi_0$  & $(R_{\rm out} - R_{\rm in})/2R_0$ & $H/R_0$ & $R_0$ & FWHM & $T_0$ & $L_{\rm UV,sub}$  & $\textcolor{black}{L_{\rm UV,sub}}/L_{\rm IR}$ \\
& deg & & deg & & & pc & days & days & $10^{46}$ erg s$^{-1}$  & \\
\hline
01003-2238 & 23 & 0 & 0 & 0.01 & 1.1 & 1.4 & 150 & -435 & 8.7 & 0.094 \\
%01572+0009 & 45 & 0.3 & 270 & 0.4 & 0.3 & 1.5 & 300 & -329 & 3.67 & 0.45 \\
01572+0009 & 20 & 0 & 0 & 0.6 & 0.2 & 1.5 & 300 & -192 & 1.63 & 1.00 \\
12071-0444 & 20 & 0 & 0 & 0.3 & 1 & 1.5 & 300 & -730 & 4.99 & 0.192 \\
12071-0444 with offset & 20 & 0.25 & 90 & 0.1 & 0.4 & 2.0 & 300 & -912 & 14.7 & 0.065 \\

\hline
\end{tabular}
\end{table*}

\begin{comment}
\begin{table}
\centering
\small
\caption{
{Grid parameters used for the torus echo model of IRAS 12071-0444.
For the centered-source model, $r_0/R_0$ and $\phi_0$ were fixed to zero, whereas for the offset-source model, these two parameters were varied over the ranges listed below. Here, $t_{\rm span}$ denotes the time span covered by the observed light curve.}
}
\label{tab:torus_grid_offset}
\begin{tabular}{ll}
\hline
Parameter & Grid values \\
\hline
$i$ [deg] 
& $0, 10, 20, 30, 40$ \\
$(R_{\rm out} - R_{\rm in})/2R_0$
& $0.1, 0.2, 0.4, 1.0$ \\
$H/R_0$ 
& $0.1, 0.2, 0.3, 0.4, 0.6, 1.0$ \\
$R_0$ [pc] 
& $1.0, 1.2, 1.3, 1.4, 1.5, 1.6, 1.7, 1.8, 1.9, 2.0$ \\
Flare FWHM [days]
& $100, 200, 300$ \\
$T_0$ 
& $-0.25t_{\rm span}$ to $0.25t_{\rm span}$ with 41 grid points \\
$r_0/R_0$
& $0.10;\ 0.15$--$0.50$ in steps of $0.05$ \\
$\phi_0$ [deg]
& $0, 45, 90, 135, 180$ \\
\hline
\end{tabular}
\end{table}
\end{comment}

%%%% 

% Any journal's BST file (e.g., apj.bst) can be used as PASJ's BST is unavailable.    
% \bibliographystyle{****}
% \bibliography{****}

\bibliographystyle{apj.bst}
\bibliography{ULIRGs}{}

\end{document}